\documentclass[aps,pre,twocolumn,superscriptaddress,showpacs]{revtex4-1}

\usepackage{graphicx}
\usepackage{natbib}
\usepackage{color}
\usepackage{amsmath}
\usepackage{float}

\begin{document}

\title{Ordering in Granular Rod Monolayers Driven Far from Thermodynamic Equilibrium}

\author{Thomas M\"uller}
\affiliation{Experimentalphysik V, Universit\"at Bayreuth, D-95440 Bayreuth, Germany}
\author{Daniel de las Heras}
\affiliation{Theoretische Physik II, Universit\"at Bayreuth, D-95440 Bayreuth, Germany}
\author{Ingo Rehberg}
\author{Kai Huang}
\email[]{kai.huang@uni-bayreuth.de}
\affiliation{Experimentalphysik V, Universit\"at Bayreuth, D-95440 Bayreuth, Germany}

\date{\today}

\begin{abstract}
The orientational order in vertically agitated granular rod monolayers is investigated experimentally and compared quantitatively with equilibrium Monte Carlo simulations and density functional theory. At sufficiently high number density, short rods form a tetratic state and long rods form a uniaxial nematic state. The length-to-width ratio at which the order changes from tetratic to uniaxial is around $7.3$ in both experiments and simulations. This agreement illustrates the universal aspects of the ordering of rod-shaped particles across equilibrium and nonequilibrium systems. Moreover, the assembly of granular rods into ordered states is found to be independent of the agitation frequency and strength, suggesting that the detailed nature of energy injection into such a nonequilibrium system does not play a crucial role.
\end{abstract}

\pacs{45.70.-n, 05.70.Ln, 64.70.M-}

\maketitle

\section{Introduction}

The ordering of anisotropic particles is a universal phenomenon appearing widely in nature, ranging from thermally driven molecules or colloids~\cite{Stephen1974, Gennes1995, Onsager1949, Vroege1992} to active particles such as bacteria colonies~\cite{Zhang2010}, actin filaments~\cite{Schaller2010, Sanchez2012}, animal groups~\cite{Buhl2006, Couzin2005, Liu2013}, and living liquid crystals~\cite{Zhou2014}. In equilibrium lyotropic systems, such as hard rods interacting only through excluded volume interactions, the transition of sufficiently anisotropic particles into various ordered states is entropy driven. The loss in rotational degrees of freedom in the ordered state is compensated by the gain in the translational ones~\cite{Onsager1949, Frenkel1999, Vroege1992}. Taking a two-dimensional system of hard rectangles as an example, a tetratic state with four-fold rotational symmetry was discovered in Monte Carlo (MC) simulations~\cite{Wojciechowski2004, Donev2006}, and studied 
theoretically with density functional theory (DFT)~\cite{velascoSPT, Martinez-Raton2009, Geng2009}. The number density and the length-to-width ratio (aspect ratio) of the particles were found to be the key parameters determining the ordered states of hard rectangles with only excluded volume interactions~\cite{velascoSPT}. Given the ubiquity of ordering transitions in nature, it is important to ask how well the existing knowledge about such transitions in equilibrium (thermal) systems can be extended to nonequilibrium (athermal) systems.

Due to the dissipative interactions between particles, agitated granular matter has been frequently used as a nonequilibrium model system for phase transitions~\cite{Jaeger1996, Ristow1997, Goetzendorfer2006, Eshuis2007, Fingerle2008, Huang2010, May2013}. Rich and often counterintuitive dynamical behavior~\cite{Boerzsoenyi2013} has been discovered for granular rods, including vortex patterns~\cite{Blair2003}, collective swirling motions~\cite{Aranson2007}, giant number fluctuations~\cite{Narayan2007, Aranson2008}, violation of the equipartition theorem~\cite{Harth2013}, and an enhanced ordering transition in an effective `thermal' bath of spherical particles~\cite{Kumar2014}. Reminiscent to equilibrium systems, ordering transitions of vertically agitated granular rods were investigated in three-dimensional (3D) and quasi-two-dimensional systems. In 3D, the aspect ratio of the rods was found to influence the ordered states of cylindrical rods~\cite{Yadav2013}. In quasi-two-dimensional systems, a bulk isotropic-uniaxial nematic (I-U) transition was observed for cylindrical rods with large aspect ratios~\cite{Galanis2006} and an effective elastic constant was characterized quantitatively~\cite{Galanis2010}. Particularly in strict monolayer systems, the shape of the rods was found to play an important role in determining the ordered states: Tetratic, nematic or smectic order was found for cylindrical rods, tapered rods or rice particles, respectively~\cite{Narayan2006}. Moreover, tetratic order was also found for tubular shaped particles and the influence of the container shape was discussed in~\cite{Sanchez2014}.

Despite all of these progresses, it is still unclear to which extent one can draw quantitative connections between systems in and out of thermodynamic equilibrium. More specifically, a quantitative comparison between the state diagram of dissipative granular rods and that of the corresponding equilibrium system is still lacking. This quantitative comparison is the purpose of the present work. Here we investigate experimentally the orientational order in monolayers of cylindrical granular rods driven far from thermodynamic equilibrium, and compare the results to MC simulations as well as DFT of the analogous equilibrium system. Focusing on the bulk region of the system, we detect both tetratic and uniaxial nematic states by varying the aspect ratio of the rods.  We demonstrate that the aspect ratio and the number density of rods are the key parameters determining the state diagram in both systems. We find a common aspect ratio that separates tetratic and uniaxial nematic states in both experiments and MC simulations. Such an agreement illustrates the universal aspects of the ordering of rod-shaped particles.

\section{Methods}

\subsection{Experiments}

\begin{figure}
\includegraphics[width=0.9\columnwidth]{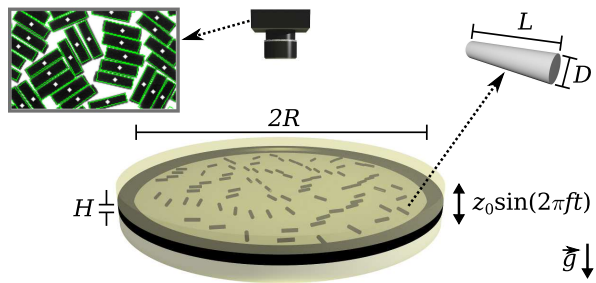}
\caption{\label{fig:setup} (Color online) Sketch of the experimental set-up. The closed cylindrical container of height $H$ and radius $R$ is driven sinusoidally against gravity with an electromagnetic shaker. The rods have a length $L$ and a diameter $D$. The embedded image shows a close view of the detected particles. }
\end{figure}

A sketch of the experimental set-up is shown in Fig.~\ref{fig:setup}. Monodisperse polyvinyl chloride (PVC) rods of diameter $D$ and length $L$, cut from welding wires of $D=3$\,mm (aspect ratio $L/D\le5$) or $1.5$\,mm ($L/D\ge5$), are confined in a cylindrical container of height $H$ and radius $R=10$\,cm. The ratio $H/D=4/3$ is chosen for both diameters to ensure a monolayer of particles; that is, no rods can cross or jump over each other. The inner surface of the container is covered with antistatic spray (Kontakt Chemie, Antistatik 100) to minimize electrostatic forces. An electromagnetic shaker (Tira TV50350) is employed to drive the sample sinusoidally against gravity with frequency $f=50$\,Hz and peak acceleration $\Gamma=4\pi^2f^2z_{\rm 0}/g$, where $z_{\rm 0}$ is the peak vibrational amplitude and $g$ is the gravitational acceleration. The acceleration is monitored with an accelerometer (Dytran 3035B2). We capture high contrast images of the rods using backlight LED illumination and a camera (IDT 
MotionScope M3) mounted above the container. The camera is synchronized with the shaker so as to capture images at a fixed phase of each vibration cycle. The images are subjected to an analysis algorithm that determines the center of mass $P_i=(x_i, y_i)$ and the orientation $\theta_i\in[0,\pi[$ of the $i$-rod with $i\in {[1,N]}$. $\theta_i$ is the angle of the main rod axis with respect to a fixed laboratory axis, and $N$ is the total number of rods in the container. The detection rate is $100$~\% for $D=3$~mm and $95$~\% for $D=1.5$~mm. 

To systematically study the collective behavior of the rods, we vary the global area fraction $\Phi_{\rm g}=\frac{NLD}{\pi R^2}$ between $\sim0.3$ and $\sim0.9$, and the aspect ratio $L/D$ between $2.0$ and $13.3$. For each $\Phi_{\rm g}$ and $L/D$, we vary the peak acceleration $\Gamma$ with a step of $1$ from $2$ to $20$ and back. The waiting time between each step is fixed at $\sim1.5$~minutes. We repeat the whole cycle at least $3$ times.

\subsection{Monte Carlo Simulations}

Correspondingly, we model the particles as two-dimensional hard rectangles of length $L$ and width $D$ that interact through excluded volume interactions. $N$ of such particles are placed in a box with dimensions $L_x$ and $L_y$ along the $x$- and $y$-axes, respectively. We use periodic boundary conditions along both axes and study the equilibrium bulk configurations by means of standard MC simulations \cite{Allen1987} in the canonical ensemble. That is, we fix the number of particles $N$ and the system area $A=L_xL_y$ (the temperature is irrelevant in hard models). The number of particles is similar to that in the experiments, $N\sim10^3$. We use simulation boxes with rectangular and square shapes. No difference has been found between the two geometries.

The simulation method is as follows. In order to equilibrate the system we start at very high area fractions, $\Phi=\frac{NLD}{A}\approx0.95$. We place the particles, with their main axes pointing in the same direction, in a rectangular lattice. Next we run $10^7$ Monte Carlo steps (MCSs). Each MCS is an attempt to move and rotate all the particles in the system. The maximum displacement $\Delta r_{\text{max}}$ and maximum rotation $\Delta\theta_{\text{max}}$ that each particle is allowed to perform in a MCS is determined such that the acceptance probability is $0.2$. Then we remove a few randomly chosen particles, recalculate $\Delta r_{\text{max}}$ and $\Delta\theta_{\text{max}}$, and start a new simulation. The number of removed particles is such that the change in area fraction is $\Delta\Phi\lesssim0.01$. In order to rule out metastable configurations related to the preparation of the initial state, we discard simulations with $\Phi\gtrsim0.8$. When the area fraction is below that limit we start the proper simulation. 
For each simulation we first run $10^6$ MCSs to equilibrate the system and then accumulate data over $10^7$ MCSs. For selected $L/D$ we have also simulated the system by increasing the number of particles, i.e., by adding particles instead of removing them. We have found no differences between both methods.

\subsection{Density functional theory}
\label{dft}
We use an Onsager-like DFT with Parsons-Lee rescaling. A similar DFT was previously used to analyze the state diagram of two-dimensional rods confined in a circular cavity~\cite{Heras2009}. We are interested in the behavior of fluid states in which the density is spatially homogeneous. Hence we can write, without loss of generality, the one body density distribution as
\begin{equation}
\rho(\vec r,\gamma)=\rho h(\gamma),
\end{equation}
where $\rho$ is the number density and $h(\gamma)$ is the orientational distribution function. Here $\gamma$ is the angle with respect to the director. $h(\gamma)$ is normalized such that
\begin{equation}
\int_0^\pi {\rm d} \gamma h(\gamma)=1.
\end{equation}
We split the free energy into two parts
\begin{equation}
F[\rho]=F_{\text{id}}[\rho]+F_{\text{ex}}[\rho],
\end{equation}
where $F_{\text{id}}$ is the ideal gas part and $F_{\text{ex}}$ is the excess part accounting for the excluded volume interactions. The ideal free energy per unit of area $A$ is given exactly by
\begin{equation}
\frac{\beta F_{\text{id}}[\rho]}{A}=\int_0^\pi {\rm d} \gamma\rho h(\gamma)\ln(\Lambda\rho h(\gamma)-1),
\end{equation}
where $\beta=1/k_{\text{B}}T$ with $k_{\text{B}}$ the Boltzmann's constant and $T$ the absolute temperature. $\Lambda$ is the (irrelevant) thermal volume that we set to one. The excess part is approximated by
\begin{equation}
\frac{F_{\text{ex}}[\rho]}{A}=\frac{\psi_{\text{ex}}(\Phi)}{4LD}\rho\int_0^\pi {\rm d} \gamma_1\int_0^\pi {\rm d} \gamma_2 h(\gamma_1)h(\gamma_2)v_{\text{exc}}(\gamma_{12}).\label{eqex}
\end{equation}
$v_{\text{exc}}(\gamma_{12})$ is the excluded area between two rectangles with relative orientation $\gamma_{12}$:
\begin{equation}
v_{\text{exc}}(\gamma_{12})=(L^2+D^2)\lvert\sin\gamma_{12}\rvert+2LD(1+\lvert\cos\gamma_{12}\rvert),
\end{equation}
and $\psi_{\text{ex}}(\Phi)$ is the excess free energy per particle of a reference system of hard disks at the same area fraction as our system of hard rectangles. The diameter of the disks is selected such that both disks and rectangles have the same area. Following Baus and Colot~\cite{Baus1987} we approximate $\psi_{\text{ex}}$ by:
\begin{equation}
\beta\psi_{\text{ex}}(\Phi)=(c_2+1)\frac{\Phi}{1-\Phi}+(c_2-1)\ln(1-\Phi),
\end{equation}
with $c_2=7/3-4\sqrt3/\pi\approx0.1280$. Eq. (\ref{eqex}) recovers the Onsager approximation in the low density limit.

Finally, the grand potential is given by
\begin{equation}
\Omega[\rho]/A=F[\rho]/A-\mu\rho,
\end{equation}
with $\mu$ the chemical potential. We minimize $\Omega$ with respect to $\rho$ and $h(\gamma)$ in order to find the equilibrium states. We use a standard conjugated gradient method to minimize the functional. We use a truncated Fourier expansion to describe $h(\gamma)$. We truncate the expansion such that the absolute value of the last coefficient in the expansion is smaller than $10^{-7}$.


\section{\label{subsecIII} Results and Discussion}

This section is organized as follows: We first introduce the ordered states observed in experiments and MC simulations in section~\ref{sec:os}. In section~\ref{sec:exp}, we analyze the influence of the container walls and the driving conditions in the experiments. Finally in section~\ref{sec:cmp}, we quantify the ordering transition threshold for various aspect ratios and compare the state diagrams obtained experimentally, via MC simulations and with DFT.

\subsection{\label{sec:os}Ordered states}

\begin{figure}
\includegraphics[width=0.6\columnwidth]{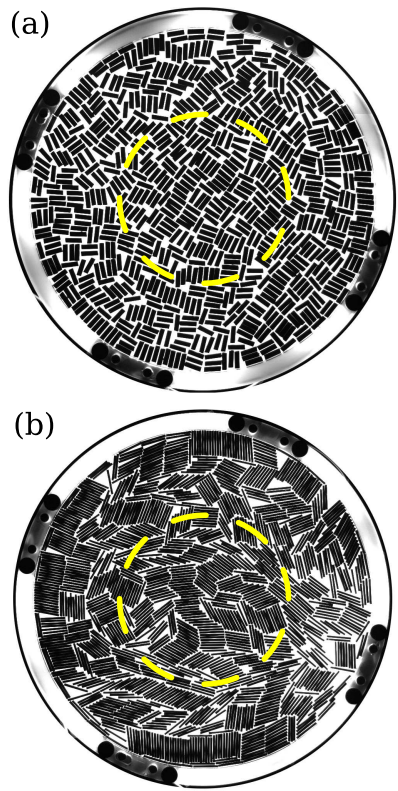}
\caption{\label{fig:snapshot} (Color online) Raw experimental images (topview) showing typical configurations of rods with aspect ratio $L/D=3.3$~(a) and  $L/D=10.0$~(b) at high global area fractions. The yellow (light gray) dashed circle indicates the region of interest.}
\end{figure}

Figure~\ref{fig:snapshot} shows typical snapshots of the ordered states obtained experimentally. Short rods~(a) tend to develop tetratic order with two alignment directions perpendicular to each other. Long rods~(b) form uniaxial nematic order with only one preferred alignment direction. In both cases, the container promotes either homeotropic (perpendicular) or planar (parallel) anchoring of the rods close to the boundary. To minimize the boundary effects, we consider only those particles located in the central region of the container, as marked in Fig.~\ref{fig:snapshot}. A quantitative justification of this region of interest (ROI) will be given in section~\ref{sec:exp}. Sometimes during the experiments, especially at low global area fractions, we observe regions with very low number density of rods (almost empty regions). As we are interested in the bulk behavior, we discard those configurations in which the ``empty regions" and the ROI overlap.

\begin{figure}
\includegraphics[width=\columnwidth]{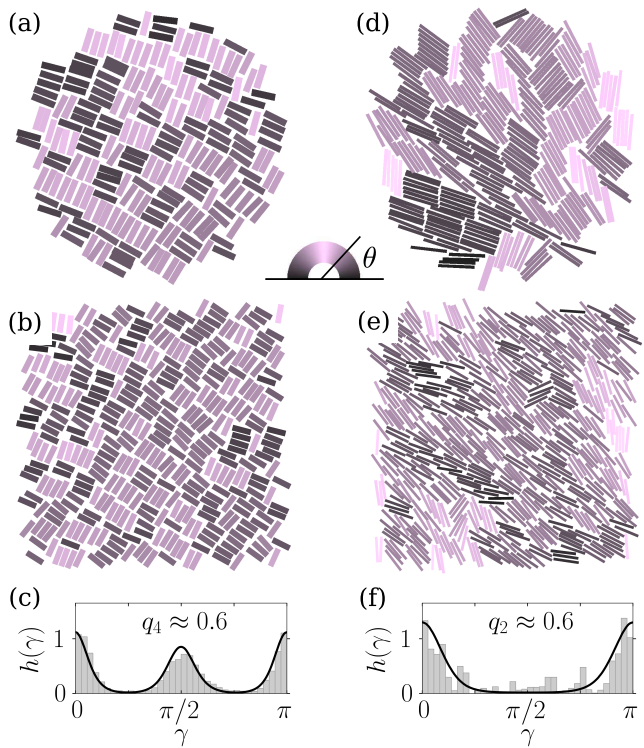}
\caption{\label{fig:sa} (Color online) Typical snapshots of tetratic (left column, $L/D=3.3$) and uniaxial nematic (right column, $L/D=10.0$) states. (a) and (d) are reconstructed from the positions and orientations of the particles detected in the center region of the container. (b) and (e) are from MC simulations with periodic boundary conditions. The particles are color coded according to their orientations. (c) and (f) show the orientational distribution functions $h(\gamma)$ of the particles in experiments (gray bars) and simulations (solid line). $\gamma$ is the angle with respect to the director.}
\end{figure}

Figure~\ref{fig:sa} shows a direct comparison of the ordered states obtained in both experiments and MC simulations. The color coded particle configurations are reconstructed from granular rods in the ROI (upper panels) and from MC simulations (middle panels) with periodic boundary conditions. In the tetratic state with fourfold rotational symmetry (left column), the orientational distribution function $h(\gamma)$, where $\gamma$ is the angle with respect to the director $\hat{n}$, has two peaks at $\gamma=0$ and $\gamma=\pi/2$ (c). In contrast, in the uniaxial nematic state (right column), the elongated particles are oriented on average along the director, yielding only one peak at $\gamma=0$ (f). The director $\hat{n}$ is calculated as the normalized eigenvector of the largest eigenvalue of the tensorial order parameter $Q_{\alpha\beta}=\langle 2w_{\alpha,i} w_{\beta,i} - \delta_{\alpha\beta}\rangle$. Here $w_{\alpha,i}$ is the $\alpha$th Cartesian coordinate of the unit vector $\hat{w}_i=(\cos \theta_i,\,\sin \theta_i)$, $\delta_{\alpha\beta}$ is the Kronecker delta, and $\langle ... \rangle$ 
denotes an average over the rods~\cite{Cuesta1990, Heras2014}. To quantify the orientational order we measure
\begin{equation}
\label{eq:op}
q_k=\int_0^\pi {\rm d}\gamma h(\gamma)\cos(k\gamma),\;\;k=\{2,4\},
\end{equation}
where $q_2$ and $q_4$ are the uniaxial and tetratic order parameters, respectively. In an isotropic state (no orientational order) $q_2$ and $q_4$ vanish. In a uniaxial nematic state $q_2>0$ and $q_4>0$. Finally in a tetratic state $q_2=0$ and $q_4>0$. The states in Fig.~\ref{fig:sa} are selected such that $q_2$ and $q_4$ are comparable in both experiments and MC simulations.

\subsection{\label{sec:exp}Experiments: The influence of boundary and driving}

\begin{figure}
\includegraphics[width=0.9\columnwidth]{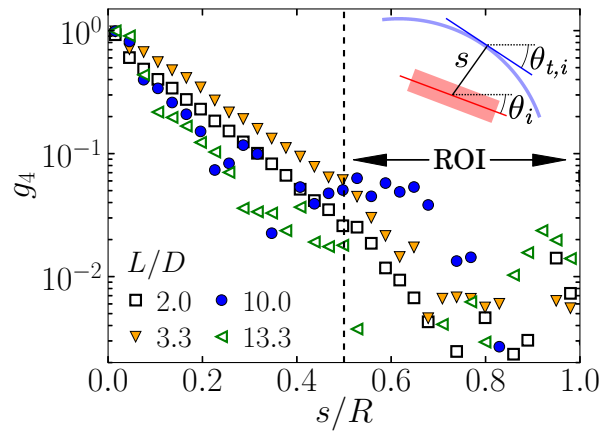}
\caption{\label{fig:wallrod_correlation} (Color online) Wall-rod angular correlation function $g_{\rm 4}$ as a function of the rescaled distance $s/R$ to the wall for various $L/D$. A sketch with various definitions is shown in the inset. The data are obtained through an average over all $\Gamma$, global area fractions $\Phi_{\rm g}$ and experimental runs. The typical error for $g_{\rm 4}$ ($\sim5\times10^{-3}$) is comparable to the size of the symbols for $s/R<0.5$. Note the logarithmic scale on the y-axis.}
\end{figure}

Experiments~\cite{Galanis2006, Narayan2006, Galanis2010a, Kudrolli2008} and MC simulations~\cite{Heras2014} have shown that the container induces a preferential alignment of the particles close to the wall. In order to facilitate the investigation in the bulk, we first need to characterize such an influence quantitatively.

Following the ideas in~\cite{Galanis2006}, we calculate the wall-rod angular correlation function $g_4(s)=\langle\cos[4(\theta_{t,i}-\theta_i(s))]\rangle$, where $s$ is the shortest distance from the rod center to the container wall, the angle $\theta_{t,i}$ quantifies the tangential direction of the corresponding point on the wall (see inset in Fig.~\ref{fig:wallrod_correlation}), and $\langle...\rangle$ denotes an average over all the particles at a distance $s$. Either homeotropic or planar alignment of the particles with respect to the wall results in $g_4 \sim 1$. In Fig.~\ref{fig:wallrod_correlation}, $g_4$ is presented as a function of the rescaled distance to the wall $s/R$ with a binning width of $0.03R$. For all aspect ratios investigated, $g_4$ decays exponentially with $s/R$. To minimize the influence of the wall, we consider only those particles with $s/R>0.5$ to be in the ROI. In this region, $g_4$ is always smaller than $0.06$ and remains in a range comparable to the experimental uncertainties. We characterize the state of the system by measuring the area fraction $\Phi$ and $h(\gamma)$ in circular regions with radius $3L$ inscribed in the ROI. Subsequently, we calculate $q_k(\Phi)$ from $h(\gamma)$ accumulated over all the regions that share the same $\Phi$.

\begin{figure}
\includegraphics[width = 0.85\columnwidth]{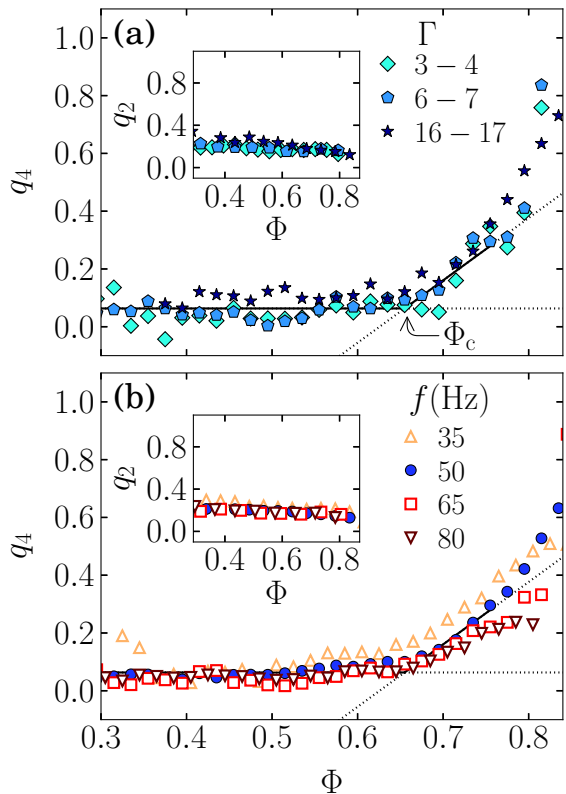}
\caption{\label{fig:varacc} (Color online) Tetratic $q_4$ and uniaxial $q_2$ order parameters in the ROI as a function of the area fraction $\Phi$ in a system of rods with aspect ratio $L/D=3.3$: (a) For three ranges of $\Gamma$ at $f=50$~Hz and (b) for four $f$ with $q_k$ accumulated over all $\Gamma$. The threshold $\Phi_{\rm c}$ is obtained through fits to the data (straight lines) accumulated over all $\Gamma$ for $f=50$~Hz (see text for details). The order parameters are not exactly zero in the isotropic state due to the finite size of the system~\cite{Cuesta1990}.}
\end{figure}

Figure~\ref{fig:varacc} shows the order parameters as a function of $\Phi$ for short rods with $L/D=3.3$. It indicates an area fraction $\Phi_{\rm c}$ above which the tetratic order parameter $q_4$ grows from its initial low value, while the uniaxial order parameter $q_2$ remains low. Such a combination of $q_2$ and $q_4$ suggests a gradual isotropic-tetratic (I-T) transition. As shown in (a), the behavior of $q_2$ and $q_4$ does not depend on the peak vibration acceleration. This is further confirmed through a comparison among data obtained for all $\Gamma$ in the range of $2\le\Gamma\le20$ and also for all aspect ratios investigated. As shown in (b), a variation of the vibration frequency $f$ from $35$~Hz to $80$~Hz for $L/D=3.3$ also yields the same behavior of $q_k(\Phi)$. 

Such agreements indicate that the details of how the rods are effectively `thermalized' in our nonequilibrium system are not essential in determining the ordering transitions, providing us the opportunity to draw connections to the corresponding equilibrium systems. Accordingly, we accumulate the data over all $\Gamma$ at $f=50$~Hz for a more accurate characterization of the transition threshold $\Phi_{\rm c}$. By fitting $q_4$ with a constant value in the isotropic region and with a straight line in the ordered state, we obtain $\Phi_{\rm c}$ as the intersection point which minimizes the standard error. Only data with sufficient statistics (i.e., error bar $<0.02$) and $q_4<0.3$ are chosen for the fits.

Moreover, the height of the container is found to play a minor role in determining the ordering transition: A variation of $H/D$ from $4/3$ to $2$ leads to the same behavior of $q_{k}$. Experiments with $H/D=2$ for $L/D=3.3$ and $L/D=10.0$ give rise to slightly lower transition thresholds $\Phi_{\rm c}$. More specifically, we find a decrease of $12\,\%$ for short rods and of $5\,\%$ for long rods, which is in both cases within the uncertainty of the fit. In addition, for a specific aspect ratio of $L/D=5.0$, the same experiments have been performed for two different rod diameters. The results agree with each other within the error bar, suggesting that the mass of the rods does not play a dominating role in the ordering transition.

\begin{figure}
\includegraphics[width=0.85\columnwidth]{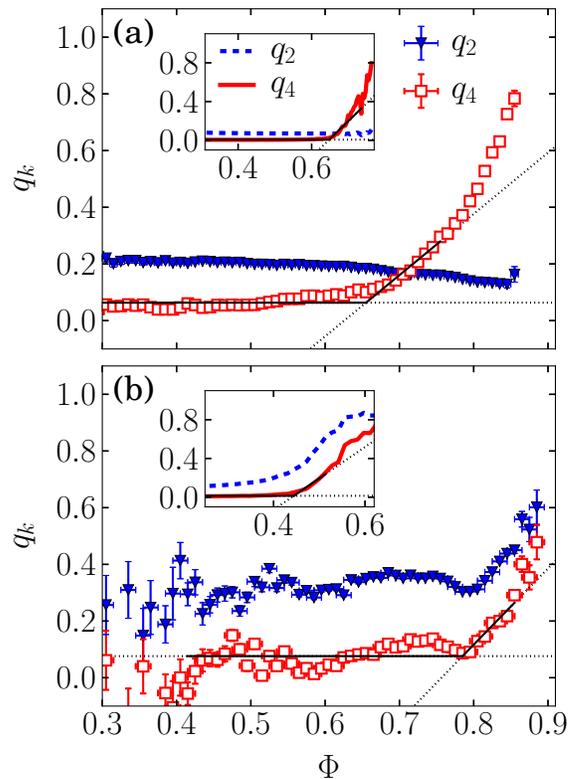}
\caption{\label{fig:order_bulk} (Color online) Tetratic and uniaxial order parameters $q_{\rm k}$ as a function of the area fraction $\Phi$ for (a) $L/D=3.3$ and (b) $L/D=10.0$. The insets show the corresponding results from MC simulations. The experimental data is an accumulation over all $\Gamma$ at $f=50$~Hz. Straight lines are linear fits to determine the threshold $\Phi_{\rm c}$.}
\end{figure}

\subsection{\label{sec:cmp}Experiments vs. simulations and DFT}

Based on the above characterizations of the boundary influence, we compare the ordering transitions of granular rods in the ROI to the corresponding thermal system. Figure~\ref{fig:order_bulk} shows the averaged order parameters obtained in both experiments and MC simulations (insets) for rods with $L/D=3.3$ (a) and $L/D=10.0$ (b). As discussed above, tetratic ordering occurs in the system of short rods. For long rods, both order parameters start to grow above $\Phi_{\rm c}$, suggesting a gradual I-U transition. Qualitatively, the agreement between experiments and MC simulations on the behavior of both tetratic $q_4$ and uniaxial $q_2$ order parameters is remarkable for both aspect ratios. Such similarities indicate that the ordering of granular rods is governed by the geometric constrain of non-overlapping rods, which is the only interaction considered in the simulations. Quantitatively, the threshold $\Phi_{\rm c}=0.66\pm 0.11$ obtained experimentally for rods with $L/D=3.3$ agrees with the one $0.65\pm 0.02$ obtained from MC simulations within the error. However, the experimentally obtained threshold $\Phi_{\rm c}=0.79\pm 0.04$ for rods with $L/D=10.0$ is larger than the one obtained for the corresponding thermal system, $0.44\pm 0.03$.

\begin{figure}
\includegraphics[width=1.00\columnwidth]{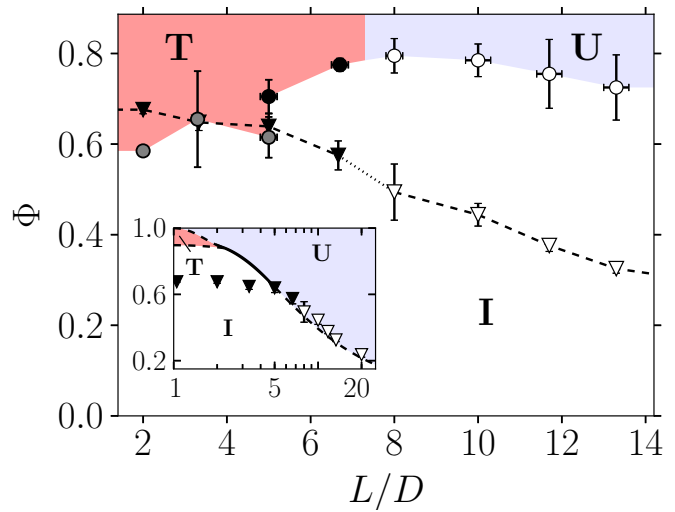}
\caption{\label{fig:phasediag} (Color online) State diagram in the plane of aspect ratio $L/D$, and area fraction $\Phi$ obtained via experiments of agitated granular rods (circles) and MC simulations of thermally driven hard rectangles (triangles). The labels denote the states: isotropic (I), uniaxial nematic (U), and tetratic (T). They are colored according to the experimental data. The diameter of the rods used in the experiments is $D=3.0$~mm for $L/D\leq5$ (gray symbols) and $D=1.5$~mm for $L/D\geq5$ (black symbols). Closed and open symbols indicate the I-T and I-U transitions, respectively. The inset shows the state diagram of equilibrium hard rods according to DFT in comparison to MC simulations in an extended region of $L/D$. Dashed lines are continuous transitions and solid lines denote first order transitions.}
\end{figure}

As $L/D$ and $\Phi$ are the key parameters determining the state of the system, we compare the experimental (nonequilibrium) results with the MC (equilibrium) simulations in a state diagram shown in Fig.~\ref{fig:phasediag}. In both systems short rods form a tetratic state and long rods a uniaxial nematic state at sufficiently high area fractions. The aspect ratio at which the ordered state changes from tetratic to uniaxial nematic agrees quantitatively. It is found to be $(L/D)_{\rm T-U}\approx7.3\pm0.7$ in both experiments and simulations~\footnote{This value represents the average between $L/D=6.6$ (uniaxial) and $L/D=8$ (tetratic).}. This result agrees with previous simulations in which a tetratic phase was found for $L/D=7$ and some evidence of uniaxial ordering for $L/D=9$ \cite{Raton2006}. The quantitative agreement of $(L/D)_{\rm T-U}$ across systems in and out of thermodynamic equilibrium illustrates the universal aspects of the ordering transitions. 

On the other hand, the threshold $\Phi_{\rm c}$ for agitated rods differs from that in MC simulations, indicating the non-universal aspects of the ordering transitions. First, the experimentally determined $\Phi_{\rm c}$ exhibit a peak around $(L/D)_{\rm T-U}$. In contrast, MC simulations show a monotonic decay with $L/D$. Second, $\Phi_{\rm c}$ measured in experiments deviates systematically from that obtained via MC simulations as $L/D$ grows. For the largest aspect ratio investigated experimentally, $L/D=13.3$, much higher area fraction is required for the uniaxial state to develop. This difference might be attributed to the following mechanisms. (i)~The strong fluctuations in the nonequilibrium steady states of granular rods may lead to temporal disorder in a system that could in principle relax into an ordered state. (ii)~Due to the dissipative rod-rod interactions, the tendency of clustering for granular rods is larger than in MC simulations, especially for large $L/D$ [compare panels (d) and (e) 
of Fig.~\ref{fig:sa}]. (iii)~Finally, the container wall may frustrate the orientational order of the agitated rods in the entire cavity. Further experiments using containers with different sizes and shapes might shed light on such a discrepancy. 

Concerning the fluctuations, it is known that the velocity distributions of agitated granular spheres are non-gaussian and exhibit exponential tails, no matter whether the particles form clusters \cite{Olafsen1998} or not \cite{Kudrolli2000}. As the dissipative nature does not depend on the shape of the particles, we expect a similar behavior in our system. This feature sets agitated granular rods apart from thermally driven liquid crystals, and triggers the question of how to define an effective `thermal' energy scale for a nonequilibrium system. Monitoring the mobility of individual granular rods with high speed photography could help to shed light on the difference between equilibrium and nonequilibrium systems found here.

In the inset of Fig.~\ref{fig:phasediag} we show the state diagram according to DFT together with the thresholds obtained from MC simulations in an extended region of $L/D$. It is similar to the one predicted by the scaled particle theory~\cite{velascoSPT}. DFT also predicts I-T transitions for small $L/D$ and I-U transitions for large $L/D$. However, the tetratic state is stable only for $L/D\lesssim2.2$, most likely because only two-body correlations are considered in the theory~\cite{Martinez-Raton2009,Raton2006}. Concerning the ordering transition threshold $\Phi_{\rm c}$, there is a good agreement between DFT and MC simulations for $L/D\gtrsim7$. For low aspect ratios, the deviations between both approaches are due to the mean field character of the theory. For $L/D<(L/D)_{\rm T-U}$, DFT predicts a T-U transition at very high area fractions. Due to the limitations in both experiments and MC simulations, the region of very high area fractions, where the T-U transition may arise, has not been 
explored.


\section{Conclusions}
To conclude, the ordering in agitated granular rod monolayers is found to be determined predominately by the aspect ratio of the rods and the area fraction, while the frequency and the strength of the agitation are not essential. It suggests that the detailed nature of energy injection into such a nonequilibrium system is not important, analogous to the role that temperature plays in equilibrium hard rod models. In comparison to previous experimental investigations on monolayer systems, we have focused on the bulk region of the container and found both tetratic and uniaxial nematic order for cylindrical rods. This enables a direct comparison to the state diagram of the corresponding equilibrium system. We find that, depending on whether the aspect ratio is smaller or larger than $\approx7.3$, a gradual isotropic-tetratic or an isotropic-uniaxial nematic transition arises as the area fraction grows, in both experiments and simulations. This agreement suggests some degree of universality for the ordering of rod shaped particles across systems in and out of thermodynamic equilibrium. Nevertheless, we have also found a qualitative difference between both systems, namely the trend of the area fraction threshold at the ordering transitions.

Further investigations will focus on characterizing the area fraction and velocity fluctuations of the system, in order to find an effective `thermal' energy scale for such a nonequilibrium system. Moreover, a comparison to molecular dynamics simulations \cite{Volfson2004} with tunable rod-rod dissipation energy could help to elucidate how fluctuations influence the ordering transition threshold.

\begin{acknowledgments}  
The authors would like to thank Wilhelm August for the preliminary work on the experimental set-up. Inspiring discussions with M. Schmidt, D. van der Meer, and C. Kr\"ulle are greatly acknowledged. TM and KH acknowledge the support from the DFG through Grant No.~HU1939/2-1. 
\end{acknowledgments}

%

\end{document}